%% file: hcomp16.tex
\def\year{2016}\relax
\documentclass[letterpaper]{article}

\usepackage{graphicx}
\usepackage{amsmath,amssymb}
\usepackage{aaai16}
\usepackage{times}
\usepackage{booktabs}
\usepackage{url}
\usepackage[usenames, dvipsnames]{color}
\definecolor{mygray}{gray}{0.6}
\definecolor{darkred}{rgb}{0.25, 0.0, 0.0}
\definecolor{darkgreen}{rgb}{0.0, 0.5, 0.0}

\newcommand{\smallsec}[1]{\paragraph{#1.}}
\newif\ifarxiv
\arxivtrue

\begin{document}

\title{Much Ado About Time: Exhaustive Annotation of Temporal Data}

\newcommand{\authSpace}{,\ \ }
\newcommand{\ispace}{\ \ \ \ }
\author{Gunnar A. Sigurdsson$^{1}$\authSpace 
Olga Russakovsky$^{1}$\authSpace
Ali Farhadi$^{2,3}$\authSpace  
Ivan Laptev$^{4}$\authSpace 
Abhinav Gupta$^{1,3}$ \\ 
 $^{1}$Carnegie Mellon University \ispace
 $^{2}$University of Washington \ispace
 $^{3}$The Allen Institute for AI \ispace
  $^{4}$INRIA \ispace
 \\ 
}

\maketitle
\begin{abstract}
Large-scale annotated datasets allow AI systems to learn from and build upon the knowledge of the crowd. Many crowdsourcing techniques have been developed for collecting image annotations. These techniques often implicitly rely on the fact that a new input image takes a negligible amount of time to perceive. In contrast, we investigate and determine the most cost-effective way of obtaining high-quality multi-label annotations for temporal data such as videos. Watching even a short 30-second video clip requires a significant time investment from a crowd worker; thus, requesting multiple annotations following a single viewing is an important cost-saving strategy. But how many questions should we ask per video? We conclude that the optimal strategy is to ask \emph{as many questions as possible} in a HIT (up to 52 binary questions after watching a 30-second video clip in our experiments).
We demonstrate that while workers may not correctly answer all questions, the cost-benefit analysis nevertheless favors consensus from multiple such cheap-yet-imperfect iterations over more complex alternatives. When compared with a one-question-per-video baseline, our method is able to achieve a $10\%$ improvement in recall ($76.7\%$ ours versus $66.7\%$ baseline) at comparable precision ($83.8\%$ ours versus $83.0\%$ baseline) in about half the annotation time ($3.8$ minutes ours compared to $7.1$ minutes baseline). We demonstrate the effectiveness of our method by collecting multi-label annotations of 157 human activities on 1,815 videos.
\end{abstract}

\section{Introduction}

Large-scale manually annotated datasets such as ImageNet~\cite{deng2009imagenet} led to revolutionary development in computer vision technology. In addition to playing a critical role in advancing computer vision, crowdsourced visual data annotation has inspired many interesting research questions: How many exemplars are necessary for the crowd to learn a new visual concept~\cite{patterson2015tropel}? How can image annotation be gamified~\cite{von2004labeling,von2006peekaboom}? How can we provide richer annotators in the form of visual attributes~\cite{patterson2014sun} or object-object interactions~\cite{krishnavisualgenome}? How can we exhaustively annotate all visual concepts present in an image~\cite{deng2014scalable}?

\begin{figure}[t]
    \centering
    \includegraphics[width=0.95\linewidth]{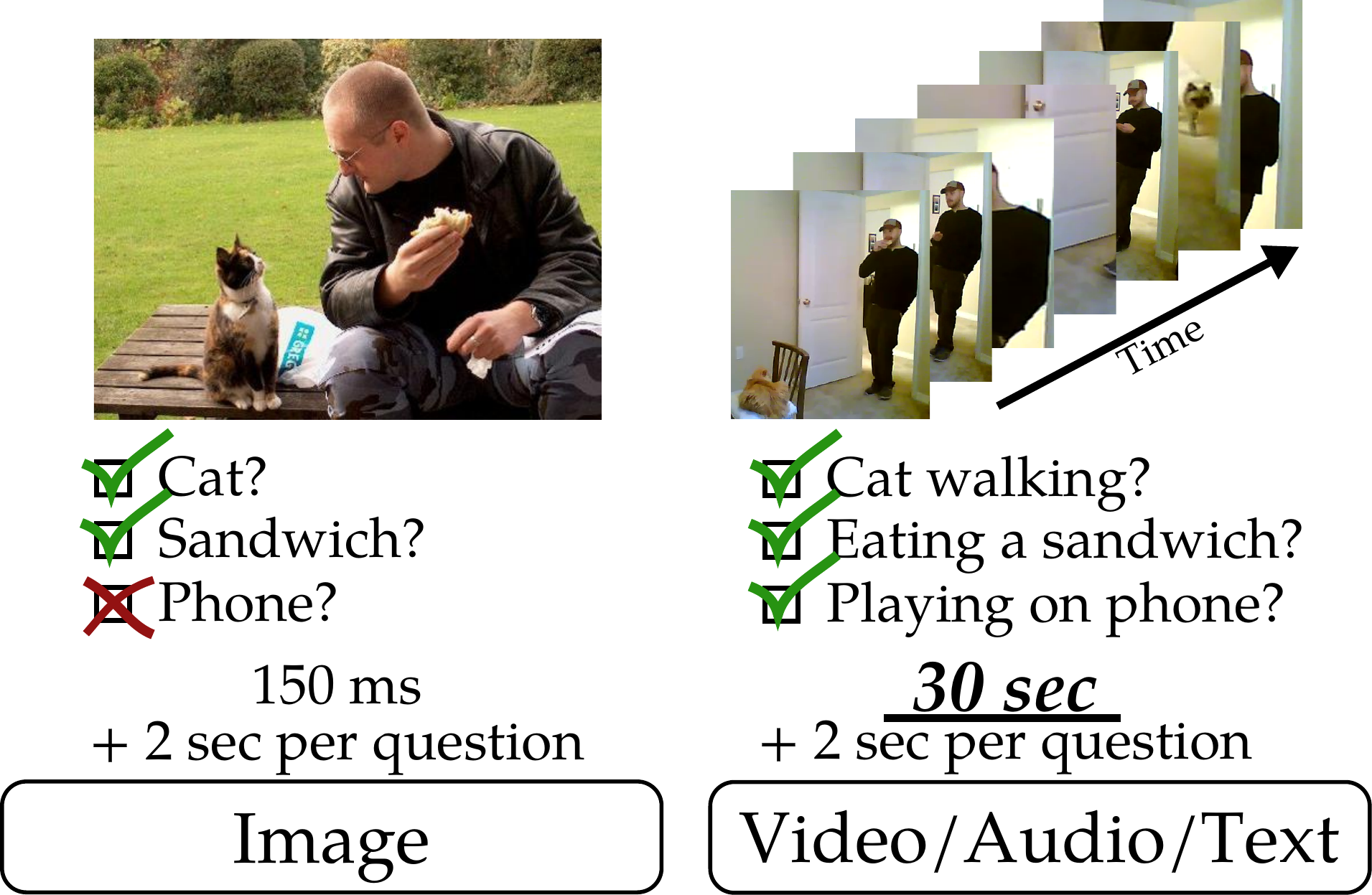}
    \caption{Time data (e.g., video) is fundamentally different from image data. In this work, we explore cost-optimal strategies for exhaustively annotating video data.}
    \label{fig:teaser}
\end{figure}

Much of the work on visual data annotation has focused on images, but many real-world applications require annotating and understanding \emph{video} rather than image data. A worker can understand an image in a few hundred milliseconds~\cite{thorpe1996speed}. Na{\"i}vely applying image annotation techniques to data that takes longer to understand, such as data involving time, is prohibitively expensive. Developing effective strategies for temporal annotation is important for multiple domains that require watching, listening, or reading: musical attributes or emotion on songs~\cite{li2003detecting}, web page categorization~\cite{ueda2002parametric}, news article topics~\cite{schapire2000boostexter}, and video activity recognition~\cite{Charades}.   

In this work, we are interested in the following annotation task illustrated in Fig.~\ref{fig:teaser}: given a video and a set of visual concepts (such as a set of objects or human actions or interesting events), label whether these concepts are present or absent in the video. Efforts such as Glance~\cite{lasecki2014glance} focus on quickly answering a question about a video by parallelizing the work across the crowd workforce in 30-second video clips. They are able to get results in near real-time, allowing for \emph{interactive} video annotation. In contrast, we are interested in annotating a large-scale video dataset where multiple questions (known apriori) need to be answered about each video. Even for a short 30-second video clip, it takes at least 15 seconds at double speed for an annotator to watch the video; thus, asking only a single question at a time is highly inefficient. Efforts such as~\cite{deng2009imagenet,bragg2013crowdsourcing} explore multi-label annotation of images but cannot be directly applied to temporal video data because of this inefficiency. 

We thus ask: how many questions should we ask workers when annotating a video? Psychology research shows that only on the order of 7 concepts can be kept in short-term memory~\cite{miller1956magical}. However, our results demonstrate asking many more questions at a time in a single Human Intelligence Task (HIT) can be significantly more efficient. In particular, we demonstrate that asking as many questions as possible, up to 52 questions at a time about a 30-second video in our experiments, provides an optimal tradeoff between accuracy and cost. When compared with a one-question-at-a-time baseline, our method achieves a $10\%$ improvement in recall ($76.7\%$ ours versus $66.7\%$ baseline) at comparable precision ($83.8\%$ ours versus $83.0\%$ baseline) in about half the annotation time ($3.8$ minutes ours compared to $7.1$ minutes baseline). We empirically verify that our conclusions hold for videos of multiple lengths, explore several strategies for reducing the cognitive load on the workers in the context of video annotation and demonstrate the effectiveness of our method by exhaustively annotating a video dataset of~\cite{Charades} enabling computer vision research into multi-label human action understanding.

The annotated data and additional details are available at:
\url{http://allenai.org/plato/charades/}.

\section{Related Work}

\smallsec{Video annotation applications}  Video understanding is important for many applications ranging from behavior studies~\cite{Coan07} to surveillance~\cite{Salisbury15} to autonomous driving~\cite{KITTI}. Large-scale annotated computer vision video datasets~\cite{THUMOS,UCF101,kuehne2011hmdb,caba2015activitynet,Yeung15} enable the development of algorithms that are able to automatically process video collections. However, the lack of large-scale multi-label video datasets makes it difficult to study the intricate interactions between objects and actions in the videos rather than focusing on recognition of one or a handful of concepts. 

\smallsec{Efficient video annotation} Video annotation is very time-consuming. Determining the absence of a concept in an image takes on the order of seconds; in contrast, determining the absence of a concept in a video takes time proportional to the length of the video. Efforts such as \cite{Yuen09,vondrick2013efficiently,Vija12} exploit temporal redundancy between frames to present cost-effective video annotation frameworks. The approaches of \cite{Vondrick11,Vija12,Fathi11} and others additionally incorporate active learning, where the annotation interfaces learns to query frames that, if annotated, would produce the largest expected change in the estimated object track. However, these methods combine human annotation with automatic computer vision techniques, which causes several problems: (1) these techniques are difficult to apply to challenging tasks such as activity recognition where computer vision models lag far behind human ability; (2) these methods are difficult to apply to scenarios where very short or rare events, such as shoplifting, may be the most crucial, and (3) the resulting hybrid annotations provide unfair testbeds for new algorithms. 

Glance~\cite{lasecki2014glance} focuses on parallelizing video annotation effort and getting an answer to a single question in real-time. Our work can be effectively combined with theirs: they parallelize annotation in 30-second video chunks, while we explore the most effective ways to obtain multiple labels simultaneously for every 30-second video.

\smallsec{Action recognition datasets} Some existing large-scale action datasets such as EventNet~\cite{EventNet} or Sports-1M~\cite{Karpathy14} rely on web tags to provide noisy video-level labels; others like THUMOS~\cite{THUMOS} or MultiTHUMOS~\cite{Yeung15} employ professional annotators rather than crowdsourcing. 

There are two recent large-scale video annotation efforts that successfully utilize crowdsourcing. The first effort is ActivityNet~\cite{heilbron2014collecting} which uses a proposal/verification framework similar to that of ImageNet~\cite{deng2009imagenet}. They define a target set of actions, query video search engines for proposal videos of those actions and then ask crowd workers to clean up the results. The second effort~\cite{Charades} entirely crowdsources the creation of a video dataset: one worker writes a video script containing a few target objects/actions, another one acts out the script and films the video, and others verify the work. In both these efforts, each video comes pre-associated with one or a handful of action labels, and workers are tasked with verifying these labels. In contrast, we're interested in the much more challenging problem of multi-label video annotation beyond the provided labels.

\smallsec{Multi-label image annotation} Increasingly more complex image annotations are provided in recent dataset~\cite{bigham2010vizwiz,COCO,krishnavisualgenome}. Multi-label image annotation has been studied by e.g.,~\cite{von2004labeling,deng2014scalable,bragg2013crowdsourcing,zhong2015regionspeak,PlateMate}. We incorporate insights from these works into our video annotation framework. We use a hierarchy of concepts to accelerate multi-label annotation following~\cite{deng2014scalable,bragg2013crowdsourcing}. Inspired by \cite{krishna2016embracing}, we explore using cheap but error-prone annotation interfaces over thorough but more expensive formulations.

\section{Method for multi-label video annotation}
\label{sec:exhaustive}

We are given a collection of $M$ videos and a set of $N$ target labels: for example, a list of target object classes, e.g., ``cat,'' ``table,'' or ``tree,'' or a list of human actions, e.g., ``reading a book'' or ``running.'' The goal is to obtain $M \times N$ binary labels, corresponding to the presence or absence of each of the $N$ target concepts in each of the $M$ videos. These labels can then be used for a variety of applications from training computer vision models~\cite{Charades} to studying human behavior~\cite{Coan07}.

We are particularly interested in situations where the label space $N$ is large: $N=157$ in our experiments. As a result, the key challenge is that workers are not able to remember all $N$ questions at the same time; however every time a worker is required to watch a video of length $L$ during annotation, they have to invest an additional $L$ seconds of annotation time. We focus on video annotation but our findings may be applicable to any media (e.g., audio, text) where a non-trivial amount of time $L$ is required to process each input. 

\subsection{Multiple question strategy}

Our strategy is to ask all $N$ target questions at the same time about each video, even if $N$ is much higher than the 7 concepts that people can commit to short-term memory~\cite{miller1956magical}. We randomize the order of questions and ask workers to select only the concepts that occur within the video. This naturally leads to lower recall $r$ than if we ask only a handful of questions that the workers would be more likely to read carefully. However, there are two advantages.

\smallsec{Advantage \#1: Low annotation times} Since only one worker has to watch the video instead of asking $N$ different workers to annotate one label each, this recall $r$ is obtained with relatively little time investment $t$. This makes it a highly effective strategy combined with consensus among multiple workers~\cite{sheshadri2013square}. Given a fixed time budget $T$, we can repeat the annotation process $\frac{T}{t}$ times with different workers. Assume the workers are independent and we count the concept as present in the image if at least one worker annotates it. Our expected recall in $T$ time is:
\begin{equation}
\label{eq:exprecall}
\mbox{ExpectedRecall} = 1 - (1-r)^{\frac{T}{t}}
\end{equation}
since each worker will miss a concept with $1-r$ probability, and a concept won't be annotated only if all $\frac{T}{t}$ workers independently miss it. 

\smallsec{Advantage \#2: High precision} The $M \times N$ label matrix is naturally sparse since most concepts do not occur in most videos. When workers are faced with only a small handful of concepts and none of them occur in the video, they may get nervous that they are not doing the task correctly and provide erroneous positive labels. However, when they are faced with many concepts at the same time and asked to select the ones that occur in the video, they get satisfaction out of being able to annotate several target concepts and are less likely to erroneously select additional concepts.

\subsection{Practical considerations}

\begin{figure}[t]
	\centering
	\includegraphics[width=0.95\linewidth]{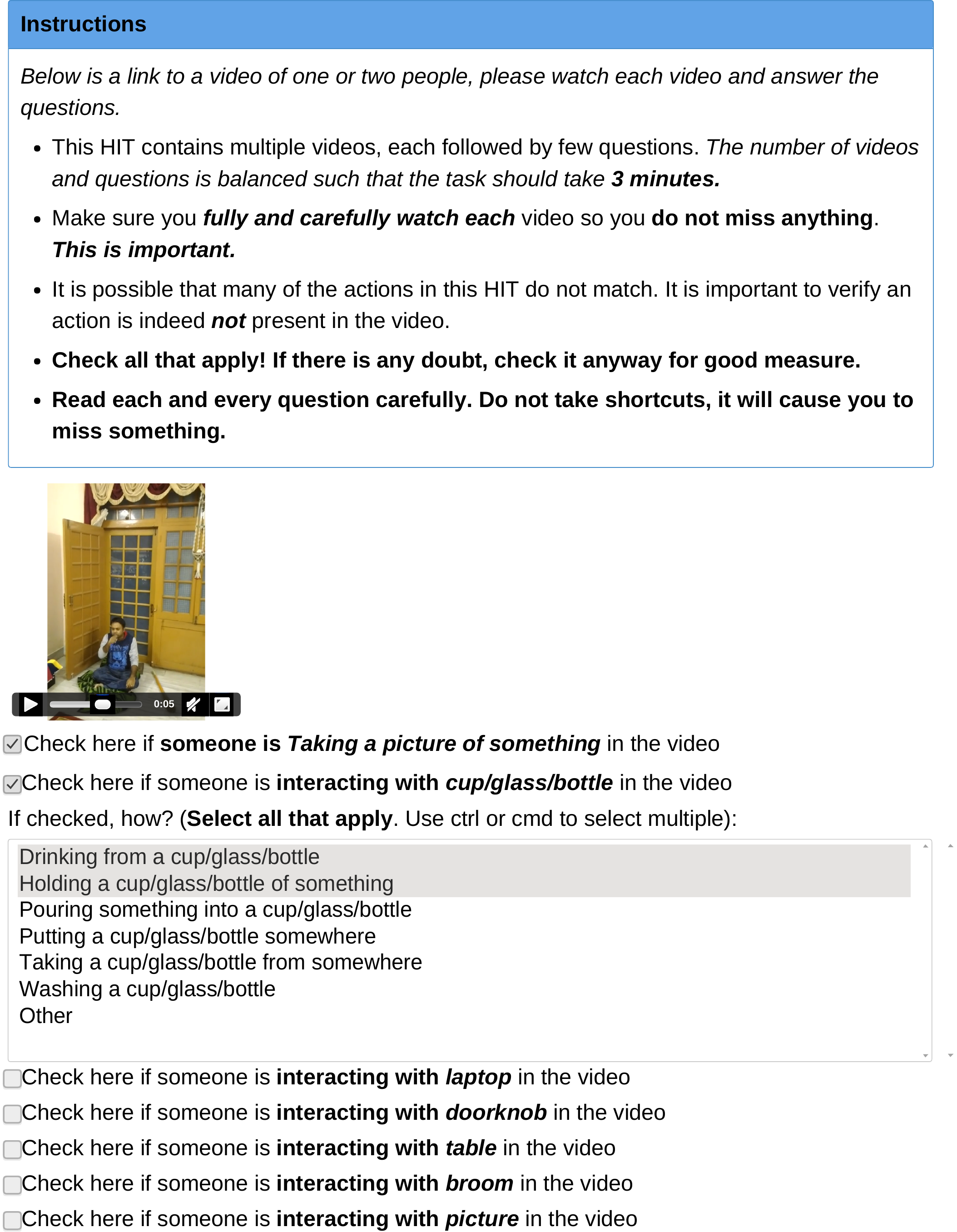}
	\caption{Our multi-question video annotation interface.}
	\label{fig:interface}
\end{figure}

In designing an effective multi-question video annotation interface shown in Fig.~\ref{fig:interface}, we incorporate insights from image annotation~\cite{deng2014scalable} to reduce the space of $N$ labels and from video annotation~\cite{lasecki2014glance} to compress the video length $L$. 

\smallsec{Semantic hierarchy} Following~\cite{deng2014scalable} we create a semantic hierarchical grouping of concepts to simplify the multi-label annotation. However, \cite{deng2014scalable} use the hierarchy differently. They ask one question at a time about a matrix of images, e.g., ``click on all images which contain an animal.'' They then ask a low-level question, e.g., ``click on all images which contain a dog,'' on a smaller matrix of images which were positive for the prior question. In contrast, we use the concept hierarchy similar to~\cite{SUN} to simplify our annotation interface on a single video. 

\smallsec{Playback speed} Videos of average length of 30 seconds are played at 2x speed following~\cite{lasecki2014glance}. In this way, worker time is not unnecessarily wasted but they are able to perceive and accurately annotate the target concepts.

\smallsec{Instructions} Workers are instructed to carefully watch each video and select all concepts that occur. Since most concepts do not occur in the video, workers are asked to only check the boxes for the ones that do occur and to ignore the others. We experimentally verify this design choice below. 

\smallsec{Time vs cost} We use human time and cost interchangeably throughout the paper. We group together multiple videos into a single HIT such that each HIT takes approximately the same time to complete: e.g., an interface with fewer questions will allow for more videos to be annotated in a single HIT. We pay a uniform amount per HIT. Thus, an interface that takes 3x less time will allow for 3x more videos per HIT which will allow for 3x fewer HITs to annotate the full dataset, which will in turn translate to a 3x reduction in cost when annotating a large-scale video dataset. 

\section{Experiments}

We begin by describing the setup used to evaluate our method, including steps taken to control for factors of variation across different crowdsourcing experiments. We then present a series of smaller-scale experiments on 100-150 videos at a time investigating (1) varying the number of questions in the annotation interface, and (2) strategies for reducing cognitive load on workers during annotation. We conclude by bringing our findings together and evaluating our large-scale multi-label video annotation pipeline.

\subsection{Data and evaluation setup}

\begin{figure}[t]
	\centering
	\includegraphics[width=1.0\linewidth]{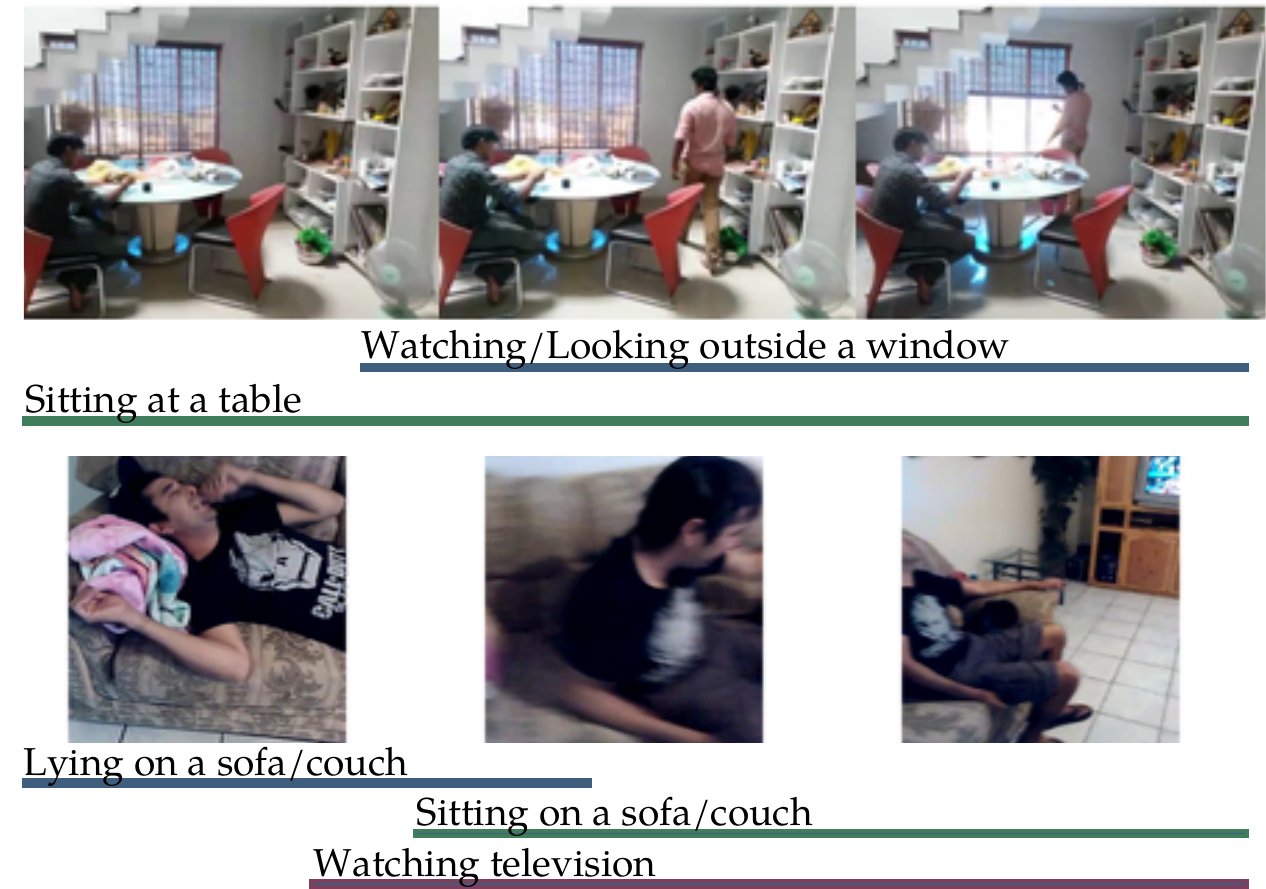}
	\caption{Examples from the video dataset of~\cite{Charades}. The videos contain complex human activities that require the annotator to carefully watch each video.}
	\label{fig:examples}
\end{figure}

We use the large-scale video dataset~\cite{Charades} with a focus on common household activities. The target labels are 157 activity classes such as \emph{Someone is running} and \emph{Putting a cup somewhere} provided with the dataset. The videos are associated with some labels apriori, similar to ImageNet~\cite{deng2009imagenet} and ActivityNet~\cite{caba2015activitynet}.  Fig.~\ref{fig:examples} shows some examples. This misses additional activities also present in the video, making it difficult to evaluate computer vision algorithms and to study interactions between different actions. We demonstrate how to cost-effectively collect exhaustive annotations for this dataset and exhaustively annotate the test set consisting of 1,815 videos. 


\smallsec{Evaluating recall} We use the originally provided action labels to evaluate the recall of our multi-label annotation algorithms. 
There were on average $3.7$ activities labeled per video in this dataset. The activities follow a long-tailed distribution: some occur in as many as $1391$ videos, others in only $33$. Each activity occurs in $42$ videos on average.

\smallsec{Evaluating precision} Precision is more difficult to evaluate since to the best of our knowledge no large-scale video dataset is annotated with hundreds of visual concepts. Annotating the videos in this dataset exhaustively in a straight-forward way is prohibitively expensive, which is exactly what we are trying to address in this work. We adopt a middle ground. After obtaining a set of candidate labels from the annotators, we perform a secondary verification step. In the verification task, workers have to annotate the temporal extent of the action in the video or specify it is not present in the video. This serves as an evaluation of the precision of multi-label annotation. In addition, this provides temporal action annotations which we also publicly released.

\smallsec{Semantic hierarchy} The 157 target human activities are grouped based on the object being interacted with to simplify the annotation interface. The annotator first sees several questions such as ``Check here if someone is interacting with a book in the video'' or ``Check here if someone is interacting with shoes in the video.'' If the annotator says \emph{yes} someone is interacting with a book, s/he will be asked to select one or more of the types of interaction: closing a book? opening a book? holding a book? putting a book somewhere?

We create 33 object groups, each group with $4.2$ activities on average. Additionally, 19 activities (such as \emph{Someone is laughing}, \emph{Someone is running somewhere}) do not belong to any group and are asked individually. 
\ifarxiv
Thus, we obtain 52 high-level questions which cover all of the label space; the exact hierarchy is provided in the Appendix.
\else
Thus, we obtain 52 high-level questions which cover all of the label space.
\fi

\subsection{Crowdsourcing setup}

During the study, $674$ workers were recruited to finish $6{,}337$ tasks on Amazon Mechanical Turk. We summarize some key crowdsourcing design decisions here.

\smallsec{Quality control} Workers were restricted to United States, with at least $98\%$ approval rate from at least 1000 tasks. We used recall, annotation time, and positive rate to flag outliers, which were manually examined and put on a blacklist. To maintain a good standing with the community all work completed without clear malice was approved, but bad workers were prohibited from accepting further work of this type.

\begin{figure}[t]
	\centering
	\includegraphics[width=0.95\linewidth]{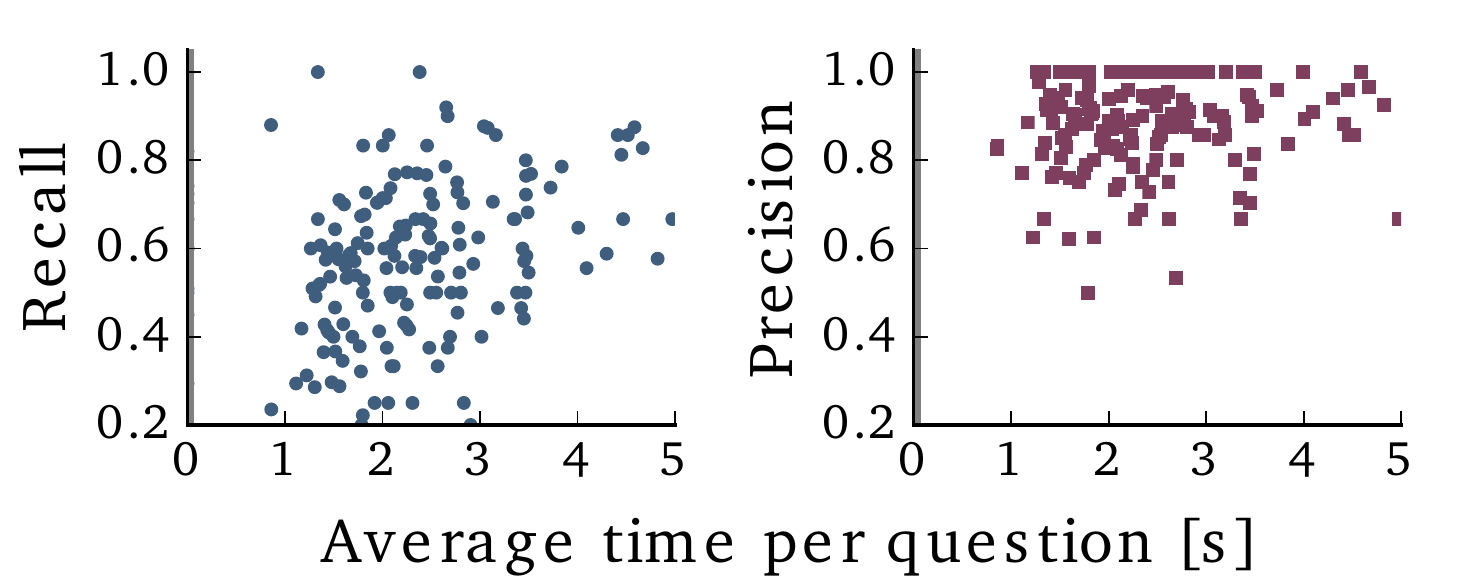}
	\caption{Workers that spend more time answering questions have marginally higher accuracy (Pearson's correlation of time with recall is $0.227$ and with precision is $0.036$). However this trend is so slight that we ignore it and instead focus on improving the annotation workflow as a whole.
	}
	\label{fig:individual}
\end{figure}

In Fig.~\ref{fig:individual} the relationship between how much time an individual worker spends on a task and quality of the annotation is presented. We can see that apart from clear outliers, there is no significant difference, and in this work we treat the worker population as following the same distribution, and focus on the time difference between different methods. 

\smallsec{Uncontrolled factors} There are many sources of variation in human studies, such as worker experience (we observed worker quality increasing as they became more familiar with our tasks) or time of day (full-time workers might primarily be available during normal business hours). We attempted to minimize such variance by deploying all candidate methods at the same time within each experiment.

\smallsec{Payment}
In order to verify our hypothesis that it is best to ask multiple questions about a video simultaneously, we need to evaluate interfaces with a varying number of questions per video. However, we want to maintain as much consistency as possible outside of the factor we're studying. We use a single type of HIT where workers are provided with $V$ videos and $Q$ questions for each video using the interface of Fig.~\ref{fig:interface}. When we increase the number of questions $Q$ per video, we decrease the number of videos $V$ to keep the expected annotation effort consistent within the HIT. 

\begin{figure}[t]
    \centering
    \includegraphics[width=0.95\linewidth]{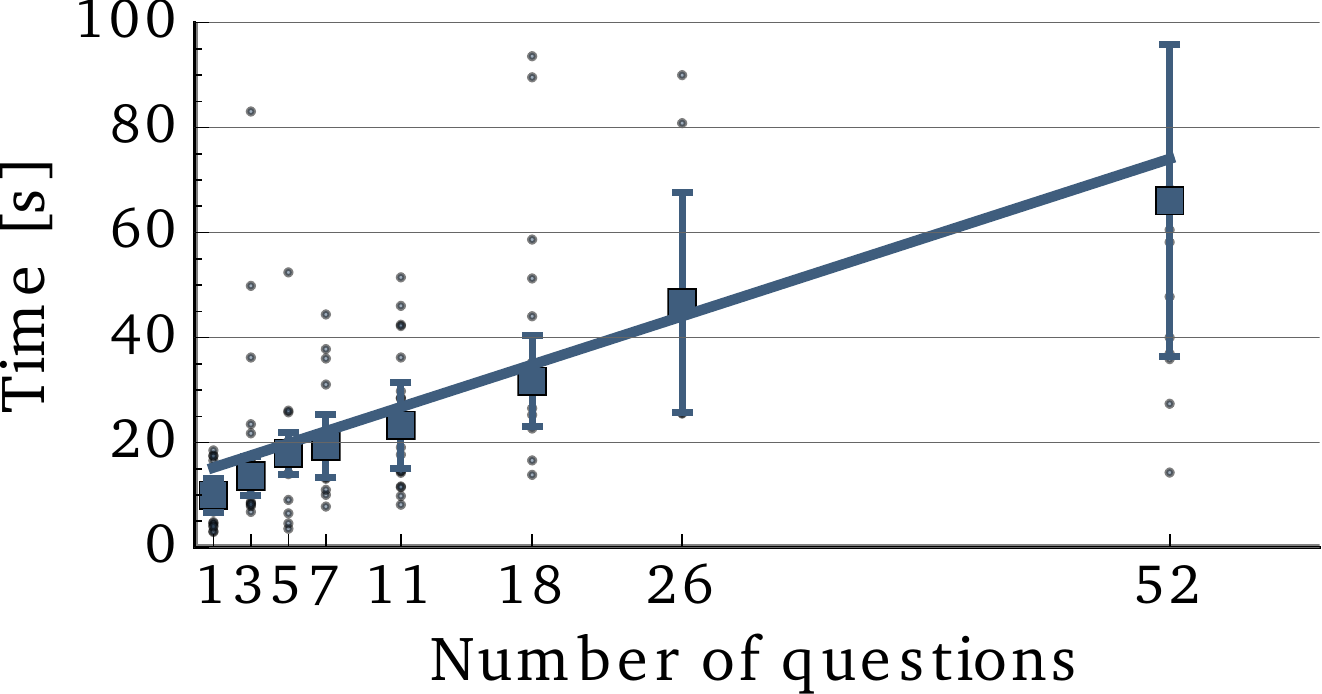}
    \caption{The relationship between number of questions in the interface and the amount of time it takes. We use it to maintain a consistent amount of annotation effort across HITs while varying the number of questions in the interface. Error bars correspond to 1 standard deviation.}
    \label{fig:ui_cost}
\end{figure}

To do this, we ran some preliminary experiments and analyzed the average amount of time it takes to label a video in our dataset with $Q$ questions. Fig.~\ref{fig:ui_cost} shows the relationship between number of questions $Q$ and time. The least-squares line of best fit to this data is 
\begin{equation}
\label{eq:time}
T = 14.1 + 1.15Q
\end{equation}
Thus it takes an average of $14.1$ seconds to watch a video and $1.15$ seconds to answer each question. This is consistent with our expectations: our average video is $30.1$ seconds long played at double speed, and binary questions take on the order of 1-2 seconds to answer~\cite{krishnavisualgenome}.

We varied the number of videos in each HIT using Eqn.~\ref{eq:time} to target about 150 seconds of expected annotation effort. We paid $\$0.40$ per HIT, amounting to about $\$9.60$ per hour.

\smallsec{Multiple question interface} We report results on annotating the 157 activities using the 52-question semantic hierarchy.\footnote{We additionally verified that all conclusions hold if we are interested in only the 52 high-level activities as well.}  Our method solicits labels for all 52 questions and corresponding sub-questions in the same interface as shown in Fig.~\ref{fig:interface}. When evaluating interfaces with a smaller number of questions $k$, we partition the 52 questions into $\frac{52}{k}$ subsets randomly. Multiple workers then annotate the video across $\frac{52}{k}$ tasks, and we accumulate the results.\footnote{Some of the questions take longer than others, and thus some subsets may take longer to annotate than others. However, we report cumulative results after all subsets have been annotated and thus the variations in time between the subsets is irrelevant.} An \emph{iteration} of annotation refers to a complete pass over the 52 questions for each video. We can then directly compare the annotations resulting from interfaces with different values of $k$.

\subsection{Effect of varying the number of questions}

So far we described the data, the evaluation metrics and the crowdsourcing setup. We are now ready to begin experimenting with different annotation strategies.

We begin by varying the number of questions the workers are asked after watching each video: from only 1 question per video (very time-inefficient since 52 workers have to independently watch the video) up to all 52 questions at the same time (potentially daunting for the workers). We run the annotation experiment on 140 videos, and report the time, recall and precision after one iteration of annotation, i.e., after workers answer all 52 questions about each video. 

\begin{figure}[t]
\centering
\includegraphics[width=0.95\linewidth]{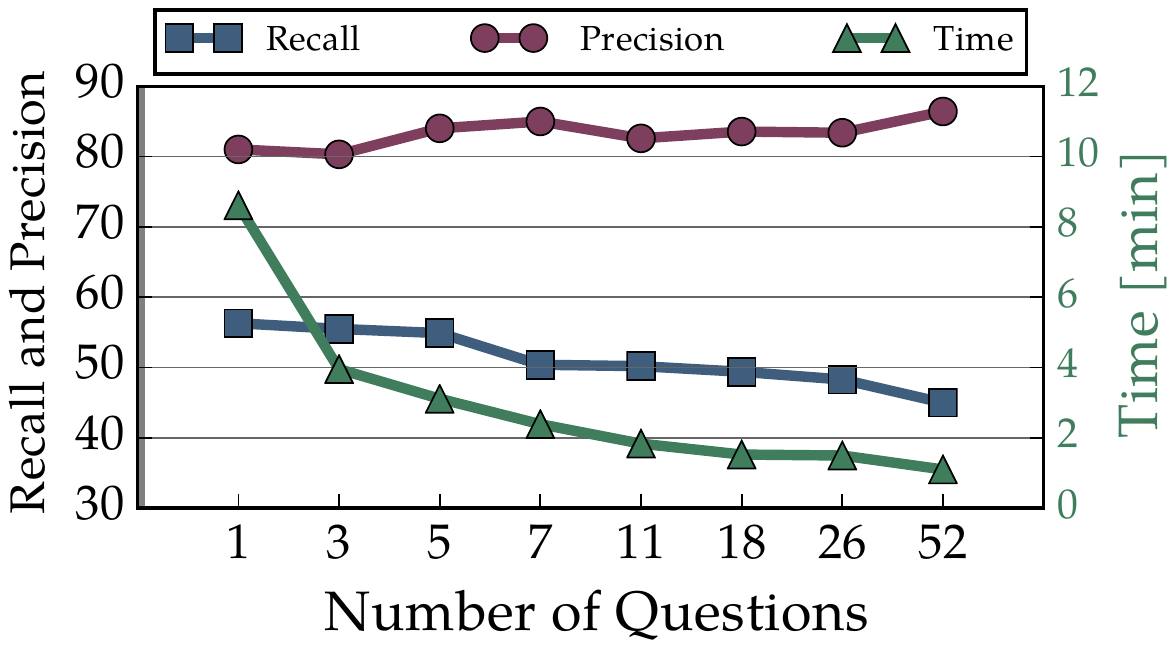}
\caption{Accuracy (\emph{left axis}) and time (\emph{right axis}) of annotation as a function of the number of questions in the interface (\emph{x-axis}). While recall is higher with fewer questions, this is at the cost of significantly higher annotation time. 
}
\label{fig:number}
\end{figure}

\smallsec{Advantages of asking multiple questions} Fig.~\ref{fig:number} demonstrates two advantages to asking multiple questions together rather than one-at-a-time. The first advantage is \emph{low annotation time}: the time for one iteration of annotation drastically decreases as the number of questions increases. Concretely, it takes $8.61$ minutes per video with the 1-question interface versus only $1.10$ minutes per video with the 52-question interface (since the time to watch the video gets amortized).

The second advantage to asking multiple questions together is \emph{increased precision} of annotation. Concretely, precision is only $81.0\%$ with the 1-question interface but rises up to $86.4\%$ with the 52-question interface. When only one question per video is asked, all answers in a HIT will likely be negative since only a handful of target activities occurs in each 30-second video. Workers report being concerned when all answers are negative. We hypothesize that as a result they may erroneously answer positively if they have any suspicions about the activity being present, which decreases the precision of annotation in the few-question interfaces.

\smallsec{Drawback of asking multiple questions} The one drawback of asking multiple questions is \emph{decreased recall}. When asked only one question per video, workers achieve $56.3\%$ recall compared to only $45.0\%$ recall when asked all 52 questions at once. This is because it is challenging to keep 52 questions in memory while watching the video or the entire video in memory while answering the questions. Interestingly, 
Fig.~\ref{fig:number} shows a sharp drop in recall beyond 5-7 questions in the interface, which is the number of concepts people can keep in short-term memory~\cite{miller1956magical}.

\begin{figure}[t]
\centering
\includegraphics[width=0.95\linewidth]{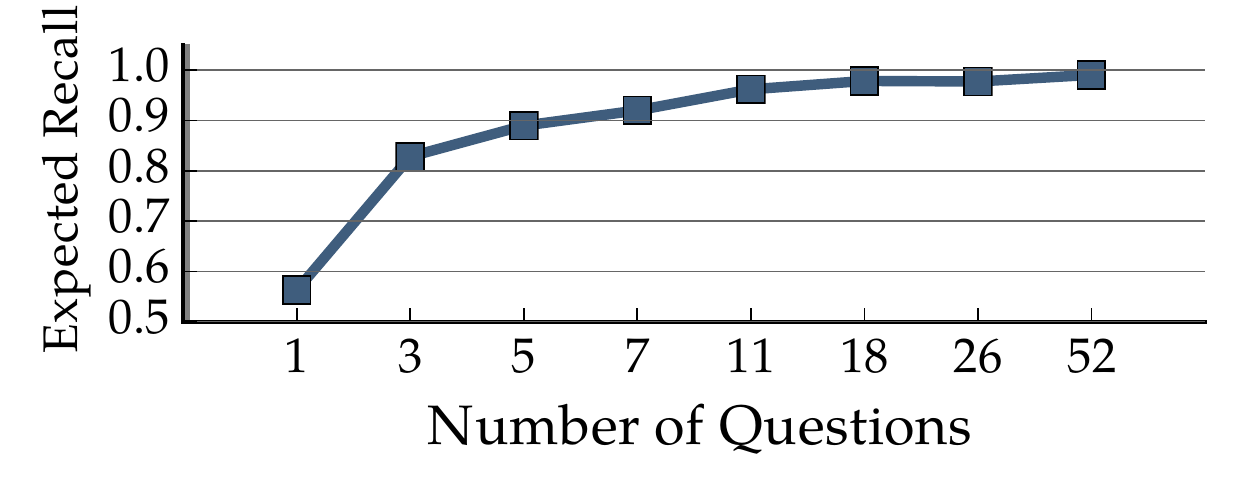}
    \caption{Expected recall given a fixed time budget (simulated using Eqn.~\ref{eq:time}) for interfaces with a varying number of questions. The budget is $8.61$ minutes per video, enough to run 1 iteration of annotation with the 1-question interface.}
    \label{fig:expected_recall}
\end{figure}

\smallsec{Fixing the drawback} Even though recall is lower when asking multiple questions about a video, it is obtained in significantly less annotation time. Given a fixed time budget, we can compute the expected recall if we were to ask multiple workers to do the annotation by referring back to Eqn.~\ref{eq:exprecall}. In particular, assume we are given $8.61$ minutes that it takes to fully annotate a video using the 1-question interface. In this amount of time, we can ask at least 7 workers to annotate it with the 52-question interface (since it only takes $1.10$ minutes per iteration). Fig.~\ref{fig:expected_recall} reports the expected recall achievable in $8.61$ minutes using the different interfaces. We conclude that the many-question interfaces are better than the few-question interfaces not only in terms of time and precision, but also in terms of recall for a fixed time budget. We will revisit this in later experiments. 

\begin{figure}[t]
    \centering
    \includegraphics[width=0.95\linewidth]{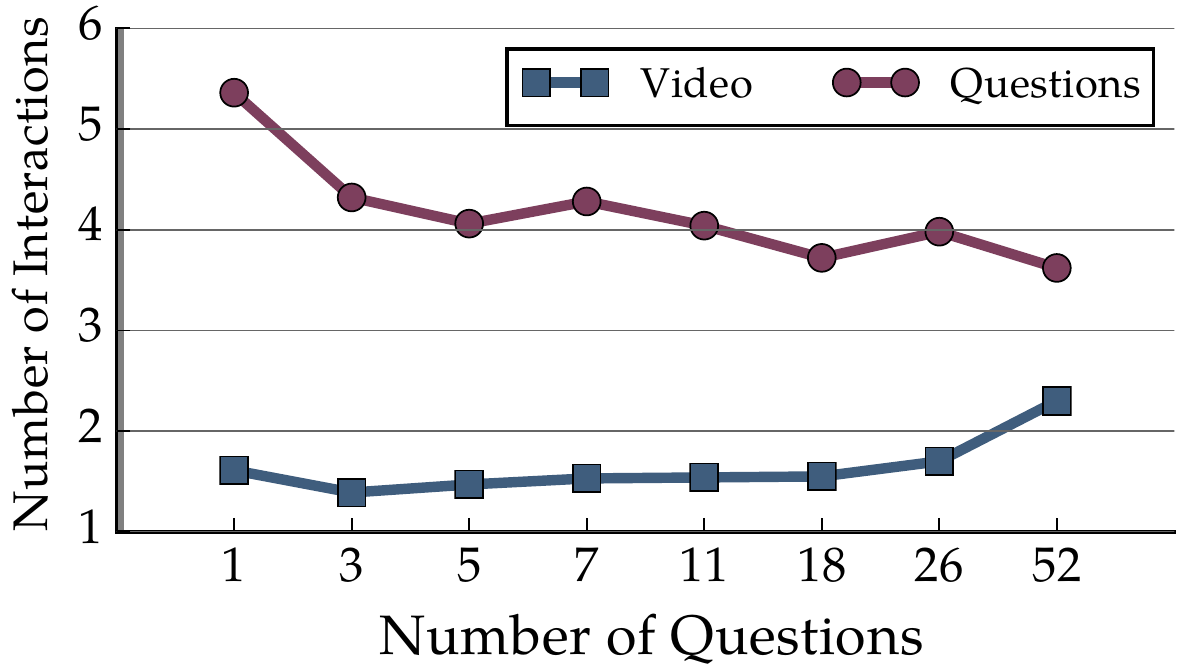}
    \caption{The number of times workers paused or synced the video (\emph{video}) and the number of questions answered affirmatively after an iteration of annotation (\emph{questions}) as a function of the number of questions in the interface. 
    } 
    \label{fig:pauseseek}
\end{figure}

\smallsec{Worker behavior} Besides quantitatively evaluating the different interfaces according to the standard metrics, it is also informative to briefly look into annotator behavior.

Fig.~\ref{fig:pauseseek} reports the number of interactions of workers with the video: i.e., the number of times they pause or seek the video. We observe that the interactions with the video generally increase with the question count, suggesting that workers may be watching the video more carefully when asked more questions. Interestingly, however, with only a single question the users seem to hurry through the video.

Fig.~\ref{fig:pauseseek} additionally reports the average number of questions answered affirmatively by the workers after an iteration of annotation. As the number of questions in the interface increases, the average number of affirmative answers after 52 questions have been answered decreases from $5.36$ to $3.62$. We hypothesize that when multiple questions are presented to the workers simultaneously, they feel satisfied once they are able to answer a handful of them positively; when faced with only a small number of questions, they feel increased pressure to find more positive answers. This contributes to both the increase in recall and the drop in precision.

\smallsec{Worker feedback} Finally, we asked workers to report their enjoyment of the task on a scale of 1 (lowest) to 7 (highest). Average enjoyment ranged from 5.0 to 5.3 across the different interfaces, indicating that workers were equally pleased with both few-question and many-question tasks.\footnote{In our preliminary experiments we did not use Eqn.~\ref{eq:time} to control for the amount of work within each HIT; worker enjoyment was then strongly inversely correlated with the amount of work.}

\subsection{Targeting the UI for different number of questions}

So far we investigated the effect that number of questions have on the accuracy and efficiency of annotation, while keeping all other factors constant. However, using the same user interface and annotation workflow for different numbers of questions may not be optimal. For example workers tend to worry when asked too many negative questions in a row in an interface with a few questions, or may not read all questions in detail in an interface with many questions.

In this section, we use the 3-question interface for the \emph{few-questions} setting, and the 26-question interface for the \emph{many-questions} setting. We run a series of experiments investigating strategies for improving the UI. We discover two strategies for improving the few-questions interface and conclude that our many-questions interface is optimal.

\smallsec{Positive bias} When using the few-questions interface, most answers within a HIT are expected to be negative since most target activities are not present in the videos. This has two undesirable effects: (1) workers may start paying less attention, and (2) workers may get nervous and provide erroneous positive answers, lowering the annotation precision.

To overcome this, we duplicate questions known to be positive and inject them such that approximately $33\%$ of the questions are expected to be positive. This forces the workers to pay closer attention and be more active in the annotation; on the downside, this increases the number of questions per annotation from 52 to 78 including the duplicates. 

In an experiment on 150 videos, injecting such positive bias into the few-questions interface improves on all three metrics: recall, precision and time of annotation. Recall increases from $53.2\%$ to $57.9\%$ with positive bias,\footnote{To maintain a fair comparison, answers to duplicate questions are ignored during evaluation. Thus the time it takes to answer them is also ignored when computing annotation time per iteration.} precision increases slightly from $79.0\%$ to $81.3\%$ with positive bias, and time for an iteration of annotation drops from $4.6$ minutes to $3.6$ minutes, likely because workers trust their work more and thus are able to annotate faster. Workers also report slightly higher enjoyment: on a scale of 1 (lowest) to 7 (highest), they report 5.8 enjoyment of the task with positive bias versus 5.5 without. We incorporate positive bias into the few-question interface in future experiments. 

\smallsec{Grouping} Prior work such as~\cite{deng2009imagenet}  demonstrated that asking about the same visual concepts across multiple images reduces the cognitive load on workers and increases annotation accuracy. In our second experiment, we apply the same intuition to videos: we randomly group  questions together and make sure that all questions are the same for all videos within a single HIT. Residual question not part of the groups, and groups too small to fill a whole task were discarded, but each question was presented both in the context of grouping and not, for a fair comparison.


In the few-questions interface, grouping improves the precision and the time of annotation, albeit at a slight reduction in recall. Specifically, in an experiment on 100 videos, precision increases from $77.7\%$ to $81.4\%$ when grouping is added. Annotation time per iteration drops from $5.9$ minutes to $5.1$ minutes with grouping; however, recall also drops from $70.4\%$ to $67.2\%$ with grouping. To determine if the drop in recall is a concern, we refer back to Eqn.~\ref{eq:exprecall} to compute the expected recall for a fixed time budget. In $5.9$ minutes (enough for one iteration without grouping), we expect a recall of $72.3\%$ with grouping, higher than $70.4\%$ recall without. Thus, we conclude that grouping is strictly beneficial in the few-question setting as hypothesized, and we use it in future experiments.

We also investigated the effect of grouping in the many-question interface, but concluded it's unhelpful. Recall with grouping is $55.2\%$, much lower than $62.0\%$ without grouping. Even though annotation time is faster ($1.4$ minutes per iteration with grouping compared to $1.6$ minutes per iteration without), this is not enough to compensate for the drop in recall: the expected recall given a budget of $1.6$ minutes of annotation is still only $61.2\%$ with grouping compared to $62.0\%$ without. Further, precision is also lower with grouping: $79.0\%$ with grouping compared to $80.2\%$ without. We hypothesize that this is because workers are not able to remember all 26 questions anyway, so grouping only provides a false sense of security (as evidenced by the speedup in annotation time). We do not use grouping in the multi-question interface in future experiments.

Note that grouping had no effect on worker enjoyment. On a scale of 1 (lowest) to 7 (highest), workers reported 5.30 enjoyment with grouping and 5.24 without. We believe this is because while grouping makes the task easier, the workers are also less engaged since it is more repetitive.

\smallsec{Video summary} Having discovered two strategies for improving the few-question interface (positive bias and grouping), we turn our attention to strategies targetting the multi-question setup. The main challenge in this setting is that workers may be overwhelmed by the number of questions and may not read them all carefully. 

To better simulate a scenario where the worker has to pay careful attention to the video, we add an additional prompt to the many-questions interface. In an experiment on 100 videos, workers were asked to ``please describe with approximately 20 words what the person/people are doing in the video.'' This adds on average $36$ seconds per iteration, yielding $2.1$ minutes of annotation time with the additional prompt versus $1.5$ without. However, the extra time does not translate to noticeable benefits in annotation accuracy: recall drops slightly to $53.2\%$ with the prompt compared to $54.2\%$ without, although precision increases slightly to $88.3\%$ with the prompt compared to $87.1\%$ without. We conclude that adding the prompt has no significant impact on the accuracy of annotation despite a $1.4$x increase in annotation time. 

\smallsec{Forced responses} The final investigation into improving the many-questions interface is asking workers to actively select a yes/no response to every question rather than simply checking a box only if an action is present. Intuitively this forces the workers to pay attention to every question. However, this again produces no improvements in accuracy, indicating that workers are already working hard to provide the most accurate responses and are only confused by the additional forced responses.

Concretely, we experimented on 100 videos and observed a drop in recall to $55.7\%$ with the forced responses compared to $63.3\%$ without as well as a drop in precision to $84.6\%$ with forced responses compared to $88.8\%$ without. Further, annotation time increases to $2.2$ minutes per video with forced responses versus $1.6$ minutes without. Thus forcing workers to read every question is in fact appears harmful: it is better for them to focus on watching the video and only skim the questions.

\smallsec{Conclusions} We thoroughly examined the annotation interface in the few-questions and many-questions setting. We discover that positive bias and grouping are effective strategies for improving the few-questions UI, and incorporate them in future experiments. For the many-questions setting, simply randomizing the questions and allowing the workers to select the actions that appear in the video is shown to be more effective than any other baseline.

\subsection{Multi-iteration video annotation}

So far we established that (1) the many-questions interface provides a more effective accuracy to annotation cost tradeoff on expectation than the few-questions interface when all other factors are kept the same, (2) the few-questions interface can be further improved by the addition of positive bias and grouping, and (3) the many-questions interface we proposed is optimal as is. In this section we bring all these findings together and conclusively demonstrate that our many-question annotation strategy is strictly better than the few-questions alternatives for practical video annotation.

\begin{figure}[t]
\centering
\includegraphics[width=0.95\linewidth]{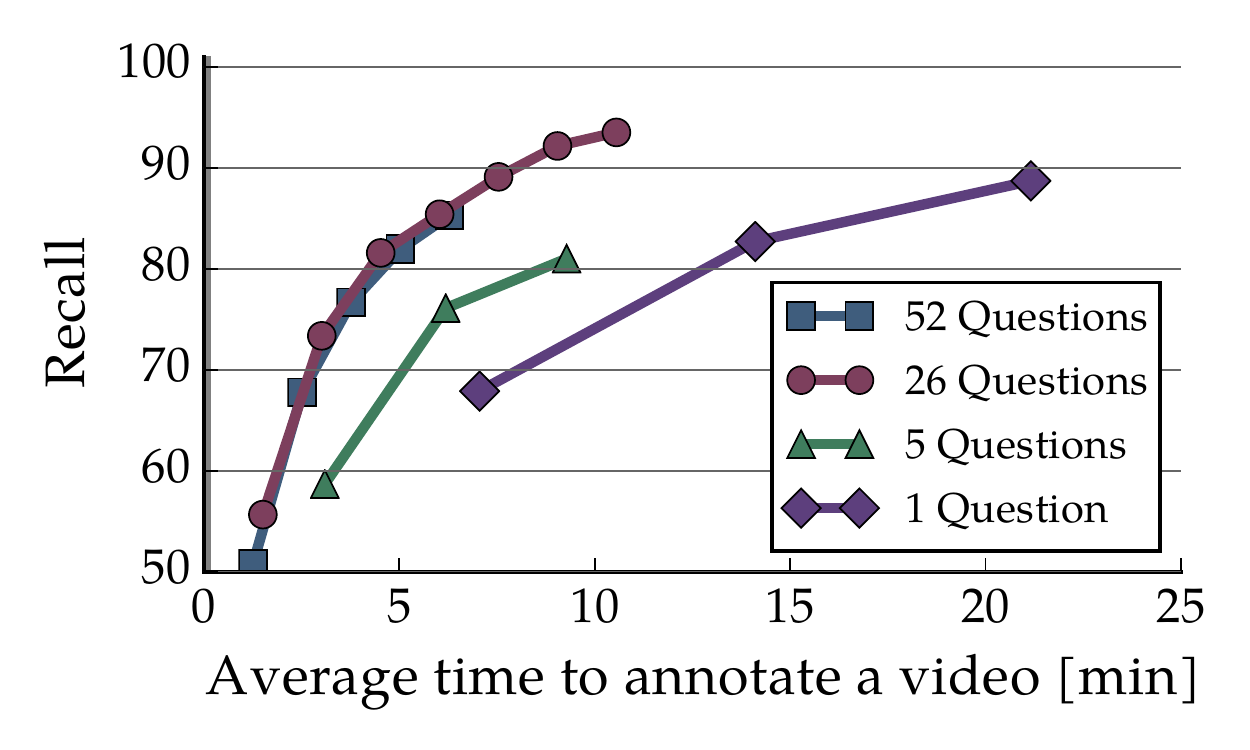}
\includegraphics[width=0.95\linewidth]{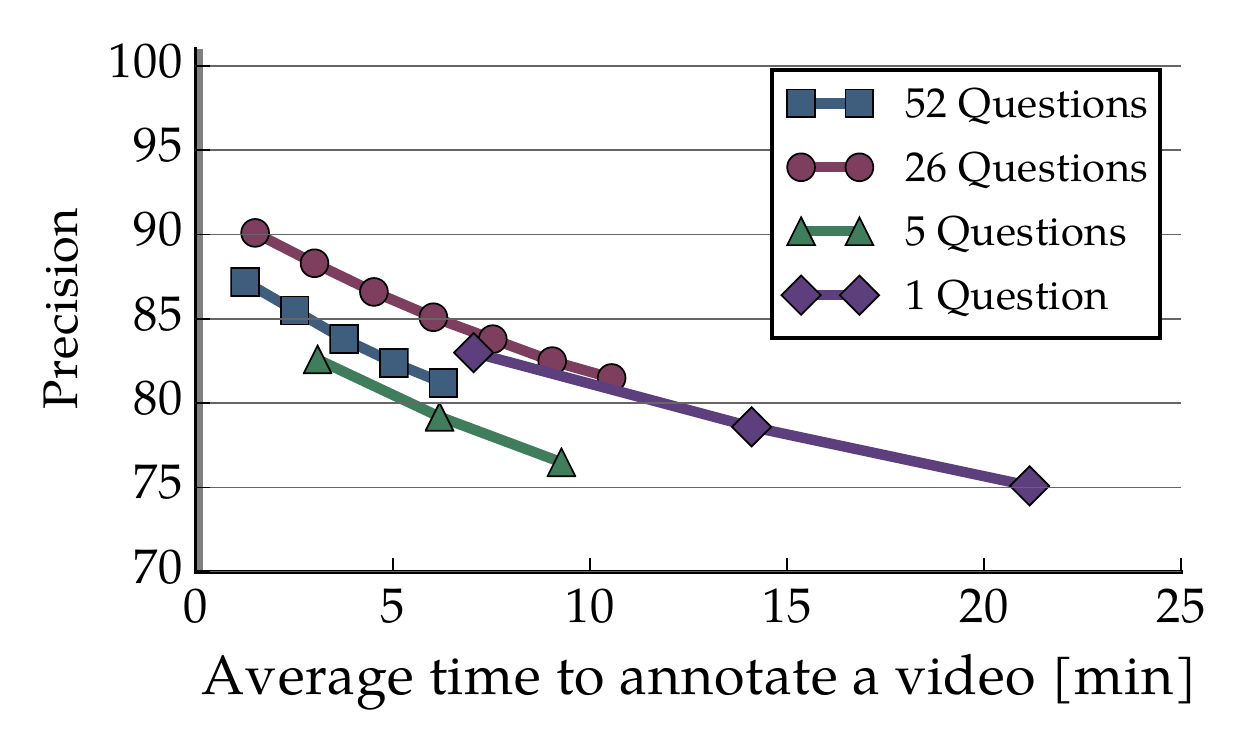}
\caption{Recall (\emph{top}) and precision (\emph{bottom}) with multiple iterations of annotation. Each square represents one iteration. We can see that since each annotation iteration with the 52-question interface is much cheaper, it quickly matches the performance of the more time-costly alternatives.}
\label{fig:consensus}
\end{figure}

\smallsec{Advantages of asking multiple questions} In previous sections we computed the expected recall across multiple iterations of annotations for a fixed time budget to compare different methods; here, we report the results in practice. We run multiple iterations of annotation and consider a label positive if at least one worker marks it as such. Thus, recall steadily increases with the number of iterations while precision may drop as more false positives are added.
 
Fig.~\ref{fig:consensus} reports recall and precision as a function of annotation time. For the few-question interfaces (5-questions and 1-question) we include the positive bias and grouping strategies found helpful above. Nevertheless, we observe a clear advantage of the multi-question methods.

For example, given $7.1$ minutes required to annotate a video with the 1-question interface, we are able to run two iterations with the 5-question interface (taking up $6.2$ minutes), and five iterations with 52-questions (taking up $6.3$ minutes). With this annotation budget, the 52-question interface obtains a recall of $85.3\%$, which is $10.5\%$ higher than the $74.8\%$ recall with 5-questions and $18.6\%$ higher than the $66.7\%$ recall with 1-question. Further, the 52-question interface obtains precision of $81.2\%$, which is $6.6\%$ higher than the $74.6\%$ precision with 5-questions and slightly lower by $1.8\%$ than the $83.0\%$ precision with the 1-question interface.

In another example, in about half the annotation time ($3.8$ versus $7.1$ minutes) we achieve a $10\%$ improvement in recall ($76.7\%$ with three iterations of 52-questions versus $66.7\%$ with one iteration of 1-question) at comparable precision ($83.8\%$ with 52-questions versus $83.0\%$ with 1-question). The improvement in recall is statistically significant at $0.01$ level using a one-tailed unequal variance t-test.

We conclude that simultaneously asking multiple questions per video, as many as 26 or even 52, is significantly more effective than asking only a handful of questions. When comparing the 26-question and  52-question interfaces in Fig.~\ref{fig:consensus}, the results are remarkably similar: recall per unit time is almost identical, although precision is slightly (statistically insignificantly) higher for 26-questions. Thus while there are diminishing returns with asking more questions in the same interface, asking ``too many'' questions per video does not appear to be harmful to annotation quality. Studying the exact point at which the number of questions becomes overwhelming for workers is important future work.  

\begin{figure}[t]
	\centering
	\includegraphics[width=0.95\linewidth]{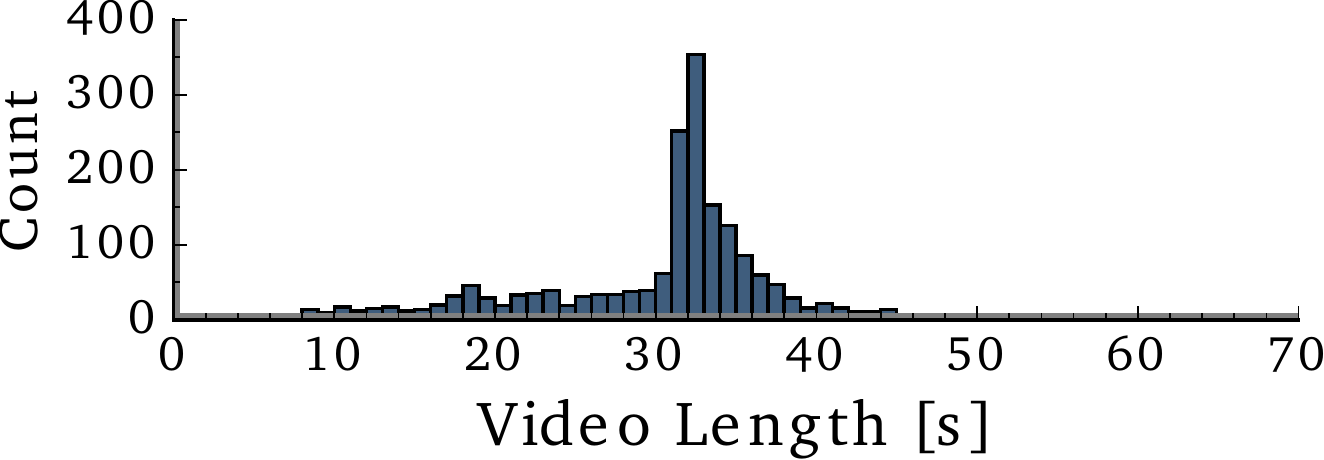}
	\caption{Statistics from the dataset. Histogram of the lengths of the videos, where we can see that the videos have various lengths enabling analysis based on content length.}
	\label{fig:lengths}
\end{figure}

\smallsec{Effect of video length} We investigate whether these conclusions hold for different video lengths -- for example, an image is just a zero-length video, so would our conclusions still apply? Our dataset contains videos of varying length as shown in Fig.~\ref{fig:lengths} and we group the videos into three groups: 0-20 seconds, 20-40 seconds and 40-60 seconds long.

\begin{figure}[!t]
	\centering
	\includegraphics[width=0.95\linewidth]{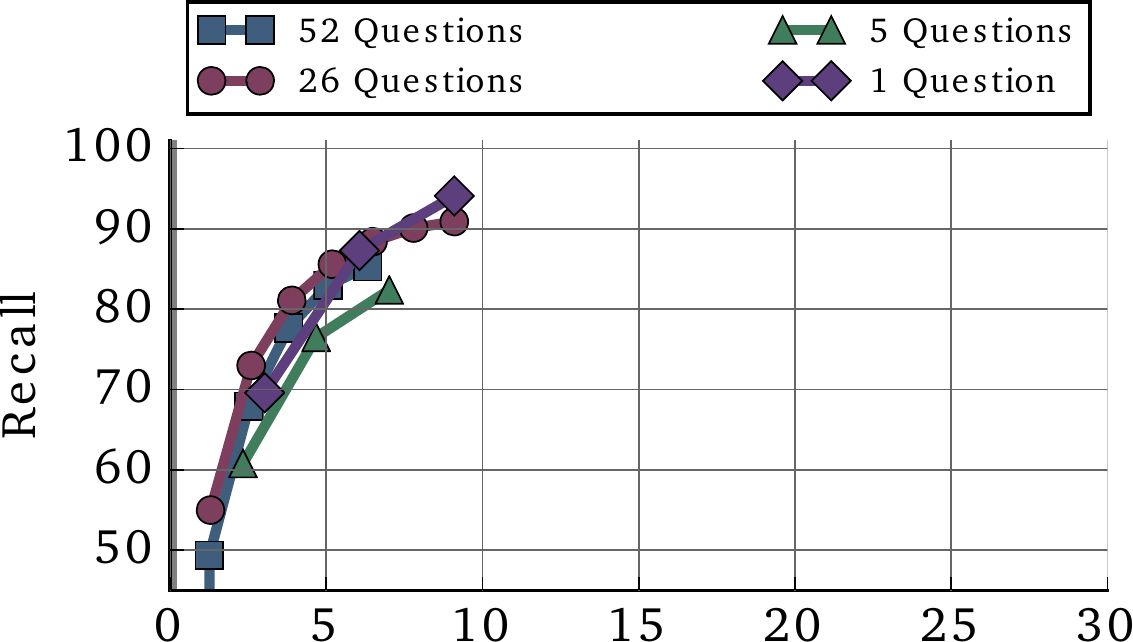}
	
	\includegraphics[width=0.95\linewidth]{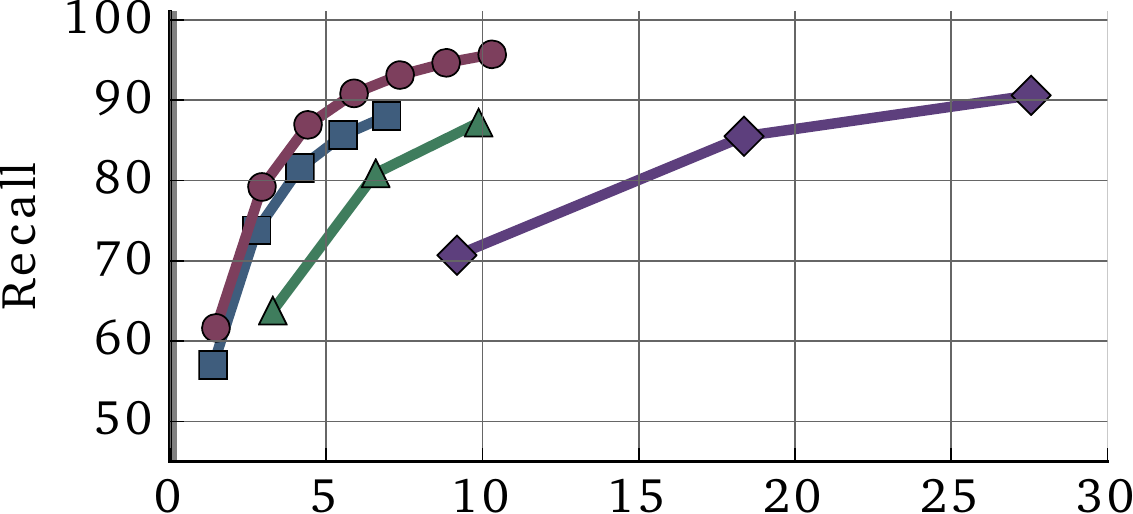}
	
	\includegraphics[width=0.95\linewidth]{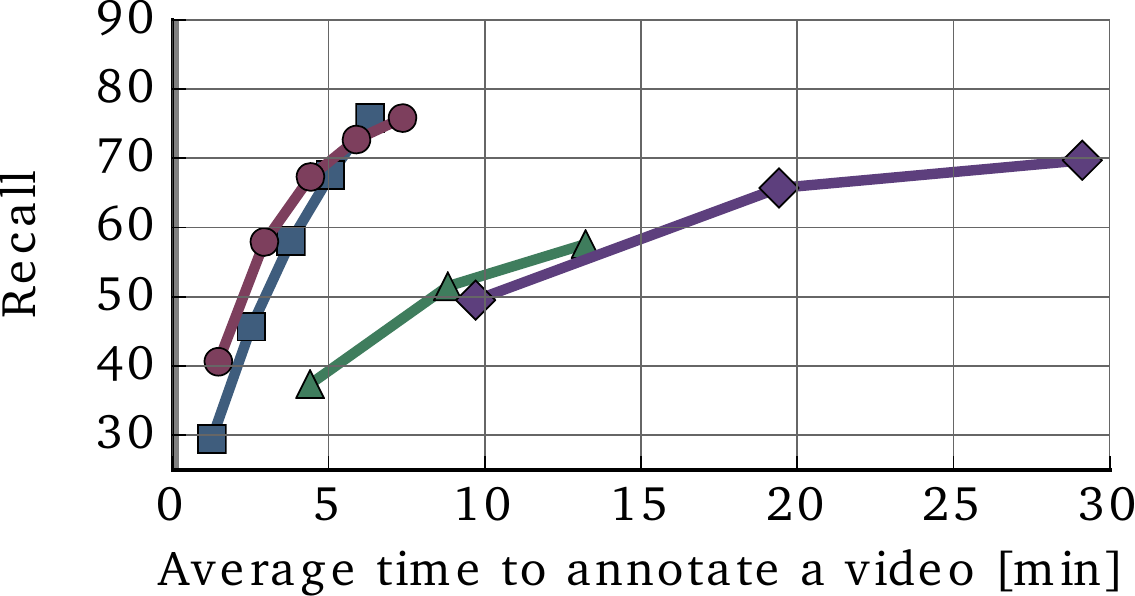}
	\caption{Breakdown of Fig.~\ref{fig:consensus} for different video lengths: \emph{(top)} 0-20 second  videos, \emph{(middle)} 20-40 second  videos, \emph{(bottom)} 40-60 second  videos. The benefit of the many-question interfaces is more prominent with increased content length.}
	\label{fig:evolution}
\end{figure}

Fig.~\ref{fig:evolution} reports the recall of the different methods for each of the three groups, following the same experimental design as before. For shorter videos that require little time to process, the exact annotation interfaces make little difference. This suggests that in the case of images our method would be as effective as the standard one-question baseline. 

Importantly, as the content gets longer the benefit of our method becomes more pronounced. For example, on 40-60 second videos for a fixed annotation budget of $4.4$ minutes (enough to run one iteration with the 5-question interface), our 52-question method achieves $62.7\%$ recall compared to only $37.4\%$ with the 5-question baseline (a $25.3\%$ improvement!) and $83.1\%$ precision compared to only $79.4\%$ precision of the 5-question baseline.

\subsection{Annotated dataset}

We used our annotation strategy to collect additional annotations for the video dataset of~\cite{Charades}. This amounted to $443{,}890$ questions answered, resulting in $1{,}310{,}014$ annotations for the $1{,}815$ videos. This increased the density of annotation on the dataset from 3.7 labels per video on average (which were available apriori based on the data collection procedure) to 9.0 labels per video. When evaluating the precision of annotation we additionally collected temporal annotation of \emph{when} the actions took place in the video. This yielded $66{,}963$ action instances. We verified the quality of temporal annotations by collecting duplicate annotations on a subset of the data. Agreement among the workers for temporal annotation was $82.8\%$ using 0.1 temporal intersection-over-union overlap. 

\begin{figure}[t]
	\centering
	\includegraphics[width=0.95\linewidth]{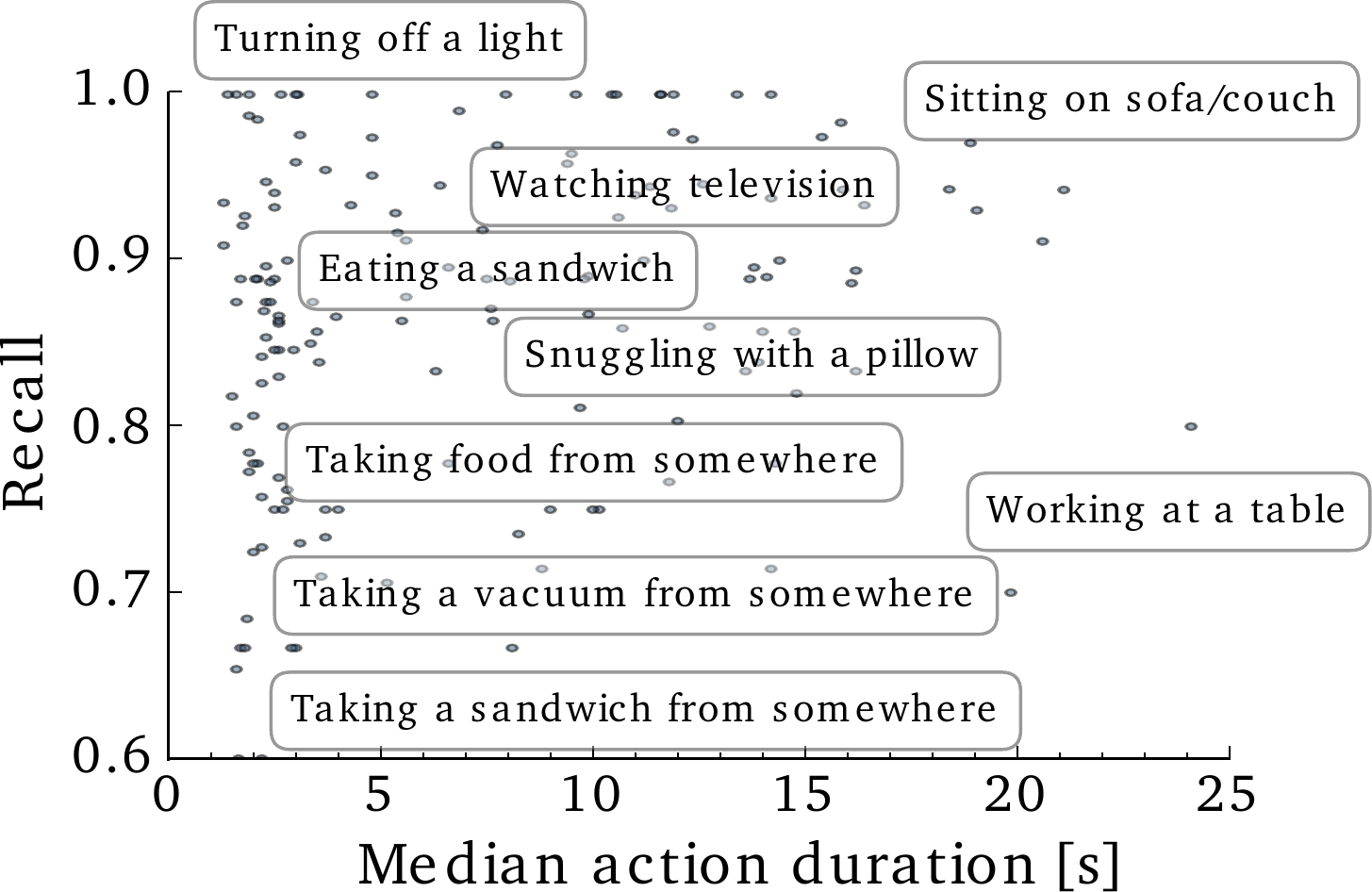}
	\caption{Annotation recall (\emph{y-axis}) as a function of the average duration in the video (\emph{x-axis}) for every one of the target 157 actions. Our method for multi-label video annotation is effective for labeling both long- and short-duration events.}
	\label{fig:analysis}
\end{figure}

Using these temporal annotations, we verify that using our method we are able to successfully annotate both actions that are long and short in the video. For every one of the 157 target actions, we compute the average (median) length of its instances in the videos as well as the recall of our annotations. Fig.~\ref{fig:analysis} plots recall as a function of action duration. As expected, recall tends to be slightly higher for actions that are longer in the video but not significantly (Pearson correlation of $0.178$). We conclude our method is effective at annotating both long and short events.

\section{Discussion \& Conclusions}

We explored the challenging problem of multi-label video annotation. In contrast to insights obtained from studying crowdsourcing of video annotation, we demonstrated that asking multiple questions simultaneously about a video provides the most effective tradeoff between annotation time and accuracy. While we observed that accuracy decreases with additional questions for each video, this drop was not sufficient to warrant the significant cost of only a few questions per video. Furthermore, we observed that the performance gap between cheap fast methods  over slow careful methods grows with increasing content length. In conclusion, our results suggest that optimal strategy of annotating data involving time is to minimize the cost in each iteration through sufficiently many questions, and simply run multiple iterations of annotation.

\section{Acknowledgements}

This work was supported by ONR MURI N00014-16-1-2007, ERC award ACTIVIA, Allen Distinguished Investigator Award, and the Allen  Institute for Artificial Intelligence. We thank Chinmay Kulkarni and G{\"u}l Varol for helpful discussions, and the Amazon Mechanical Turk workers for their time and effort.

\small
\bibliographystyle{aaai}
\bibliography{hcomp16g.bib}

\ifarxiv
\section{Appendix}

Below we present the hierarchy of concepts used in our multi-label annotation interface. The 157 human actions are organized into a hierarchy according to the object the human is interacting with. This hierarchy is used to simplify the interface. \\
\scriptsize

\input{groups.txt}
\fi

\end{document}

%% file: groups.txt
\emph{clothes}:  Holding some clothes, Putting clothes somewhere, Taking some clothes from somewhere, Throwing clothes somewhere, Tidying some clothes.

\emph{door}:  Closing a door, Fixing a door, Opening a door	.

\emph{table}:  Putting something on a table, Tidying up a table, Washing a table, Sitting at a table, Working at a table, Sitting on a table.

\emph{phone/camera}:  Holding a phone/camera, Playing with a phone/camera, Putting a phone/camera somewhere, Taking a phone/camera from somewhere, Talking on a phone/camera.

\emph{bag}:  Holding a bag, Opening a bag, Putting a bag somewhere, Taking a bag from somewhere, Throwing a bag somewhere.

\emph{book}:  Closing a book, Holding a book, Opening a book, Putting a book somewhere, Taking a book from somewhere, Throwing a book somewhere, Watching/Reading/Looking at a book.

\emph{towel}:  Holding a towel/s, Putting a towel/s somewhere, Taking a towel/s from somewhere, Throwing a towel/s somewhere, Tidying up a towel/s, Washing something with a towel.

\emph{box}:  Closing a box, Holding a box, Opening a box, Putting a box somewhere, Taking a box from somewhere, Taking something from a box, Throwing a box somewhere. 

\emph{laptop}:  Closing a laptop, Holding a laptop, Opening a laptop, Putting a laptop somewhere, Taking a laptop from somewhere, Watching a laptop or something on a laptop, Working/Playing on a laptop.

\emph{shoe/shoes}:  Holding a shoe/shoes, Putting shoes somewhere, Putting on shoe/shoes, Taking shoes from somewhere, Taking off some shoes, Throwing shoes somewhere.

\emph{chair}:  Sitting in a chair, Standing on a chair.

\emph{food}:  Holding some food, Putting some food somewhere, Taking food from somewhere, Throwing food somewhere.

\emph{sandwich}:  Eating a sandwich, Holding a sandwich, Putting a sandwich somewhere, Taking a sandwich from somewhere. 

\emph{blanket}:  Holding a blanket, Putting a blanket somewhere, Snuggling with a blanket, Taking a blanket from somewhere, Throwing a blanket somewhere, Tidying up a blanket/s. 

\emph{pillow}:  Holding a pillow, Putting a pillow somewhere, Snuggling with a pillow, Taking a pillow from somewhere, Throwing a pillow somewhere.

\emph{shelf}:  Putting something on a shelf, Tidying a shelf or something on a shelf. 

\emph{picture}:  Reaching for and grabbing a picture, Holding a picture, Laughing at a picture, Putting a picture somewhere, Watching/looking at a picture. 

\emph{window}:  Closing a window, Opening a window, Washing a window, Watching/Looking outside of a window. 

\emph{mirrow}:  Holding a mirror, Smiling in a mirror, Washing a mirror, Watching something/someone/themselves in a mirror. 

\emph{broom}:  Holding a broom, Putting a broom somewhere, Taking a broom from somewhere, Throwing a broom somewhere, Tidying up with a broom.

\emph{light}:  Fixing a light, Turning on a light, Turning off a light. 

\emph{cup/glass/bottle}:  Drinking from a cup/glass/bottle, Holding a cup/glass/bottle of something, Pouring something into a cup/glass/bottle, Putting a cup/glass/bottle somewhere, Taking a cup/glass/bottle from somewhere, Washing a cup/glass/bottle. 

\emph{closet/cabinet}:  Closing a closet/cabinet, Opening a closet/cabinet, Tidying up a closet/cabinet. 

\emph{paper/notebook}:  Someone is holding a paper/notebook, Putting their paper/notebook somewhere, Taking paper/notebook from somewhere, Working on paper/notebook. 

\emph{dish/dishes}:  Holding a dish, Putting a dish/es somewhere, Taking a dish/es from somewhere, Wash a dish/dishes. 

\emph{sofa/couch}:  Lying on a sofa/couch, Sitting on sofa/couch. 

\emph{floor}:  Lying on the floor, Sitting on the floor, Throwing something on the floor, Tidying something on the floor. 

\emph{medicine}:  Holding some medicine, Taking/consuming some medicine. 

\emph{television}:  Laughing at television, Watching television.

\emph{bed}:  Someone is awakening in bed, Lying on a bed, Sitting in a bed. 

\emph{vacuum}:  Fixing a vacuum, Holding a vacuum, Taking a vacuum from somewhere.

\emph{doorknob}:  Fixing a doorknob, Grasping onto a doorknob. 

\emph{refrigerator}:  Closing a refrigerator, Opening a refrigerator. 

\emph{misc}:  Someone is awakening somewhere, Someone is cooking something, Someone is dressing, Someone is laughing, Someone is running somewhere, Someone is going from standing to sitting, Someone is smiling, Someone is sneezing, Someone is standing up from somewhere, Someone is undressing, Someone is eating something, Washing some clothes, Smiling at a book, Making a sandwich, Taking a picture of something, Walking through a doorway, Putting groceries somewhere, Washing their hands, Fixing their hair 

%% file: hcomp16.bbl
\begin{thebibliography}{}

\bibitem[\protect\citeauthoryear{Bigham \bgroup et al\mbox.\egroup
  }{2010}]{bigham2010vizwiz}
Bigham, J.~P.; Jayant, C.; Ji, H.; Little, G.; Miller, A.; Miller, R.~C.;
  Miller, R.; et~al.
\newblock 2010.
\newblock {VizWiz}: nearly real-time answers to visual questions.
\newblock In {\em User Interface Software and Technology (UIST)}.

\bibitem[\protect\citeauthoryear{Bragg, Weld, and
  others}{2013}]{bragg2013crowdsourcing}
Bragg, J.; Weld, D.~S.; et~al.
\newblock 2013.
\newblock Crowdsourcing multi-label classification for taxonomy creation.
\newblock In {\em Human Computation and Crowdsourcing (HCOMP)}.

\bibitem[\protect\citeauthoryear{Coan and Allen}{2007}]{Coan07}
Coan, J.~A., and Allen, J. J.~B.
\newblock 2007.
\newblock {\em Handbook of Emotion Elicitation and Assessment}.
\newblock New York: Oxford University Press.

\bibitem[\protect\citeauthoryear{Deng \bgroup et al\mbox.\egroup
  }{2009}]{deng2009imagenet}
Deng, J.; Dong, W.; Socher, R.; Li, L.-J.; Li, K.; and Fei-Fei, L.
\newblock 2009.
\newblock {ImageNet}: A large-scale hierarchical image database.
\newblock In {\em Computer Vision and Pattern Recognition (CVPR)}.

\bibitem[\protect\citeauthoryear{Deng \bgroup et al\mbox.\egroup
  }{2014}]{deng2014scalable}
Deng, J.; Russakovsky, O.; Krause, J.; Bernstein, M.~S.; Berg, A.; and Fei-Fei,
  L.
\newblock 2014.
\newblock Scalable multi-label annotation.
\newblock In {\em SIGCHI Conference on Human Factors in Computing Systems}.

\bibitem[\protect\citeauthoryear{Fathi \bgroup et al\mbox.\egroup
  }{2011}]{Fathi11}
Fathi, A.; Balcan, M.; Ren, X.; and Rehg, J.
\newblock 2011.
\newblock Combining self training and active learning for video segmentation.
\newblock In {\em British Machine Vision Conference (BMVC)}.

\bibitem[\protect\citeauthoryear{Geiger, Lenz, and Urtasun}{2012}]{KITTI}
Geiger, A.; Lenz, P.; and Urtasun, R.
\newblock 2012.
\newblock Are we ready for autonomous driving? {The KITTI} vision benchmark
  suite.
\newblock In {\em Computer Vision and Pattern Recognition (CVPR)}.

\bibitem[\protect\citeauthoryear{Gorban \bgroup et al\mbox.\egroup
  }{2015}]{THUMOS}
Gorban, A.; Idrees, H.; Jiang, Y.-G.; Roshan~Zamir, A.; and Laptev.
\newblock 2015.
\newblock {THUMOS} challenge: Action recognition with a large number of
  classes.
\newblock \url{http://www.thumos.info/}.

\bibitem[\protect\citeauthoryear{Heilbron and
  Niebles}{2014}]{heilbron2014collecting}
Heilbron, F.~C., and Niebles, J.~C.
\newblock 2014.
\newblock Collecting and annotating human activities in web videos.
\newblock In {\em Proceedings of International Conference on Multimedia
  Retrieval},  377.
\newblock ACM.

\bibitem[\protect\citeauthoryear{Heilbron \bgroup et al\mbox.\egroup
  }{2015}]{caba2015activitynet}
Heilbron, F.~C.; Escorcia, V.; Ghanem, B.; and Niebles, J.~C.
\newblock 2015.
\newblock {ActivityNet}: A large-scale video benchmark for human activity
  understanding.
\newblock In {\em Computer Vision and Pattern Recognition (CVPR)}.

\bibitem[\protect\citeauthoryear{Karpathy \bgroup et al\mbox.\egroup
  }{2014}]{Karpathy14}
Karpathy, A.; Toderici, G.; Shetty, S.; Leung, T.; et~al.
\newblock 2014.
\newblock Large-scale video classification with convolutional neural networks.
\newblock In {\em Computer Vision and Pattern Recognition (CVPR)}.

\bibitem[\protect\citeauthoryear{Krishna \bgroup et al\mbox.\egroup
  }{2016a}]{krishna2016embracing}
Krishna, R.; Hata, K.; Chen, S.; Kravitz, J.; Shamma, D.~A.; et~al.
\newblock 2016a.
\newblock Embracing error to enable rapid crowdsourcing.
\newblock In {\em SIGCHI Conference on Human Factors in Computing Systems}.

\bibitem[\protect\citeauthoryear{Krishna \bgroup et al\mbox.\egroup
  }{2016b}]{krishnavisualgenome}
Krishna, R.; Zhu, Y.; Groth, O.; Johnson, J.; Hata, K.; Kravitz, J.; et~al.
\newblock 2016b.
\newblock Visual genome: Connecting language and vision using crowdsourced
  dense image annotations.
\newblock {\em CoRR} abs/1602.07332.

\bibitem[\protect\citeauthoryear{Kuehne \bgroup et al\mbox.\egroup
  }{2011}]{kuehne2011hmdb}
Kuehne, H.; Jhuang, H.; Garrote, E.; Poggio, T.; and Serre, T.
\newblock 2011.
\newblock {HMDB}: a large video database for human motion recognition.
\newblock In {\em International Conference on Computer Vision (ICCV)}.

\bibitem[\protect\citeauthoryear{Lasecki \bgroup et al\mbox.\egroup
  }{2014}]{lasecki2014glance}
Lasecki, W.~S.; Gordon, M.; Koutra, D.; Jung, M.~F.; et~al.
\newblock 2014.
\newblock Glance: Rapidly coding behavioral video with the crowd.
\newblock In {\em User Interface Software and Technology (UIST)}.

\bibitem[\protect\citeauthoryear{Li and Ogihara}{2003}]{li2003detecting}
Li, T., and Ogihara, M.
\newblock 2003.
\newblock Detecting emotion in music.
\newblock In {\em Proceedings of the fourth international conference on music
  information retrieval (ICMIR)}, volume~3,  239--240.

\bibitem[\protect\citeauthoryear{Lin \bgroup et al\mbox.\egroup }{2014}]{COCO}
Lin, T.-Y.; Maire, M.; Belongie, S.; Hays, J.; Perona, P.; Ramanan, D.; et~al.
\newblock 2014.
\newblock {Microsoft COCO: Common Objects in Context}.
\newblock In {\em European Conference on Computer Vision (ECCV)}.

\bibitem[\protect\citeauthoryear{Miller}{1956}]{miller1956magical}
Miller, G.~A.
\newblock 1956.
\newblock The magical number seven, plus or minus two: some limits on our
  capacity for processing information.
\newblock {\em Psychological review} 63(2):81.

\bibitem[\protect\citeauthoryear{Noronha \bgroup et al\mbox.\egroup
  }{2011}]{PlateMate}
Noronha, J.; Hysen, E.; Zhang, H.; and Gajos, K.~Z.
\newblock 2011.
\newblock Platemate: Crowdsourcing nutritional analysis from food photographs.
\newblock In {\em User Interface Software and Technology (UIST)}.

\bibitem[\protect\citeauthoryear{Patterson \bgroup et al\mbox.\egroup
  }{2014}]{patterson2014sun}
Patterson, G.; Xu, C.; Su, H.; and Hays, J.
\newblock 2014.
\newblock The {SUN} attribute database: Beyond categories for deeper scene
  understanding.
\newblock {\em International Journal of Computer Vision} 108(1-2).

\bibitem[\protect\citeauthoryear{Patterson \bgroup et al\mbox.\egroup
  }{2015}]{patterson2015tropel}
Patterson, G.; Van~Horn, G.; Belongie, S.; Perona, P.; and Hays, J.
\newblock 2015.
\newblock Tropel: Crowdsourcing detectors with minimal training.
\newblock In {\em Human Computation and Crowdsourcing (HCOMP)}.

\bibitem[\protect\citeauthoryear{Salisbury, Stein, and
  Ramchurn}{2015}]{Salisbury15}
Salisbury, E.; Stein, S.; and Ramchurn, S.
\newblock 2015.
\newblock Crowdar: augmenting live video with a real-time crowd.
\newblock In {\em Human Computation and Crowdsourcing (HCOMP)}.

\bibitem[\protect\citeauthoryear{Schapire and
  Singer}{2000}]{schapire2000boostexter}
Schapire, R.~E., and Singer, Y.
\newblock 2000.
\newblock Boostexter: A boosting-based system for text categorization.
\newblock {\em Machine learning} 39(2):135--168.

\bibitem[\protect\citeauthoryear{Sheshadri and
  Lease}{2013}]{sheshadri2013square}
Sheshadri, A., and Lease, M.
\newblock 2013.
\newblock Square: A benchmark for research on computing crowd consensus.
\newblock In {\em Human Computation and Crowdsourcing (HCOMP)}.

\bibitem[\protect\citeauthoryear{Sigurdsson \bgroup et al\mbox.\egroup
  }{2016}]{Charades}
Sigurdsson, G.~A.; Varol, G.; Wang, X.; et~al.
\newblock 2016.
\newblock Hollywood in homes: Crowdsourcing data collection for activity
  understanding.
\newblock In {\em European Conference on Computer Vision (ECCV)}.

\bibitem[\protect\citeauthoryear{Soomro, Roshan~Zamir, and Shah}{2012}]{UCF101}
Soomro, K.; Roshan~Zamir, A.; and Shah, M.
\newblock 2012.
\newblock {UCF101}: A dataset of 101 human actions classes from videos in the
  wild.
\newblock In {\em CRCV-TR-12-01}.

\bibitem[\protect\citeauthoryear{Thorpe \bgroup et al\mbox.\egroup
  }{1996}]{thorpe1996speed}
Thorpe, S.; Fize, D.; Marlot, C.; et~al.
\newblock 1996.
\newblock Speed of processing in the human visual system.
\newblock {\em Nature} 381(6582):520--522.

\bibitem[\protect\citeauthoryear{Ueda and Saito}{2002}]{ueda2002parametric}
Ueda, N., and Saito, K.
\newblock 2002.
\newblock Parametric mixture models for multi-labeled text.
\newblock In {\em Neural Information Processing Systems (NIPS)}.

\bibitem[\protect\citeauthoryear{Vijayanarasimhan and Grauman}{2012}]{Vija12}
Vijayanarasimhan, S., and Grauman, K.
\newblock 2012.
\newblock Active frame selection for label propagation in videos.
\newblock In {\em European Conference on Computer Vision (ECCV)}.

\bibitem[\protect\citeauthoryear{Von~Ahn and Dabbish}{2004}]{von2004labeling}
Von~Ahn, L., and Dabbish, L.
\newblock 2004.
\newblock Labeling images with a computer game.
\newblock In {\em SIGCHI conference on Human factors in computing systems}.

\bibitem[\protect\citeauthoryear{Von~Ahn, Liu, and
  Blum}{2006}]{von2006peekaboom}
Von~Ahn, L.; Liu, R.; and Blum, M.
\newblock 2006.
\newblock Peekaboom: A game for locating objects in images.
\newblock In {\em SIGCHI Conference on Human Factors in Computing Systems}.

\bibitem[\protect\citeauthoryear{Vondrick and Ramanan}{2011}]{Vondrick11}
Vondrick, C., and Ramanan, D.
\newblock 2011.
\newblock {Video Annotation and Tracking with Active Learning}.
\newblock In {\em Advances in Neural Information Processing Systems (NIPS)}.

\bibitem[\protect\citeauthoryear{Vondrick, Patterson, and
  Ramanan}{2013}]{vondrick2013efficiently}
Vondrick, C.; Patterson, D.; and Ramanan, D.
\newblock 2013.
\newblock Efficiently scaling up crowdsourced video annotation.
\newblock {\em International Journal of Computer Vision} 101(1).

\bibitem[\protect\citeauthoryear{Xiao \bgroup et al\mbox.\egroup }{2014}]{SUN}
Xiao, J.; Ehinger, K.~A.; Hays, J.; Torralba, A.; and Oliva, A.
\newblock 2014.
\newblock Sun database: Exploring a large collection of scene categories.
\newblock {\em International Journal of Computer Vision (IJCV)}.

\bibitem[\protect\citeauthoryear{Ye \bgroup et al\mbox.\egroup
  }{2015}]{EventNet}
Ye, G.; Li, Y.; Xu, H.; Liu, D.; and Chang, S.-F.
\newblock 2015.
\newblock {EventNet}: A large scale structured concept library for complex
  event detection in video.
\newblock In {\em ACM International Conference on Multimedia}.

\bibitem[\protect\citeauthoryear{Yeung \bgroup et al\mbox.\egroup
  }{2015}]{Yeung15}
Yeung, S.; Russakovsky, O.; Jin, N.; Andriluka, M.; Mori, G.; and Fei-Fei, L.
\newblock 2015.
\newblock Every moment counts: Dense detailed labeling of actions in complex
  videos.
\newblock {\em CoRR} abs/1507.05738.

\bibitem[\protect\citeauthoryear{Yuen \bgroup et al\mbox.\egroup
  }{2009}]{Yuen09}
Yuen, J.; Russell, B.; Liu, C.; and Torralba, A.
\newblock 2009.
\newblock {LabelMe} video: Building a video database with human annotations.
\newblock In {\em International Conference on Computer Vision (ICCV)}.

\bibitem[\protect\citeauthoryear{Zhong, Lasecki, and
  others}{2015}]{zhong2015regionspeak}
Zhong, Y.; Lasecki, W.~S.; et~al.
\newblock 2015.
\newblock {RegionSpeak}: Quick comprehensive spatial descriptions of complex
  images for blind users.
\newblock In {\em SIGCHI Conference on Human Factors in Computing Systems}.

\end{thebibliography}
